\documentclass[10pt, aps, twocolumn]{revtex4}
\usepackage{graphicx}
\def\vector#1{\mbox{\boldmath $#1$}}

\begin{document}

\title{Relaxation process of magnetic friction under sudden changes in velocity }

\author{Hisato \surname{Komatsu}}
\email[Email address: ]{komatsu.hisato@nims.go.jp}

\affiliation{Research Center for Advanced Measurement and Characterization, National Institute for Materials Science, Tsukuba, Ibaraki 305-0047, Japan}

\begin{abstract}

Although there have been many studies of statistical mechanical models of magnetic friction, most of these have focused on the behavior in the steady state. 
In this study, we prepare a system composed of a chain and a lattice of Ising spins that interact with each other, and investigate the relaxation of the system when the relative velocity $v$ changes suddenly. The situation where $v$ is given is realized by attaching the chain to a spring, the other end of which moves with a constant velocity $v$. Numerical simulation finds that, when the spring constant has a moderate value, the relaxation of the frictional force is divided into two processes, which are a sudden change and a slow relaxation. This behavior is also observed on regular solid surfaces, although caused by different factors than our model. More specifically, the slow relaxation process is caused by relaxation of the magnetic structure in our model, but is caused by creep deformation in regular solid surfaces.

\end{abstract}

\maketitle

\section{Introduction \label{Introduction} }

Friction is a familiar phenomenon that has been known for a long time, and there have been many studies attempting to describe its behavior\cite{PP15, BC06, KHKBC12}.
One well-known phenomenological law is the Amontons--Coulomb law. It states that frictional force $F$ is independent of relative velocity $v$. However, Coulomb himself noted that real materials violate this law slightly~\cite{PP15}. This violation was studied several decades ago, resulting in an empirical modification of the Amontons--Coulomb law known as the Dieterich--Ruina law~\cite{BC06, KHKBC12, Ruina83, Dieterich87, DK94, HBPCC94, Scholz98}. This is given by
\begin{equation}
F = F_0 + A \log v + B \log \theta ,
\label{DRlaw}
\end{equation}
\begin{equation}
\mathrm{where} \ \ \frac{d \theta}{dt} = 1 - \frac{v \theta}{D} ,
\label{DRlaw_theta}
\end{equation}
and $A, B, F_0$, and $D$ are constants. According to this law, $F$ depends on the hysteresis through Eq. (\ref{DRlaw_theta}) in the general situation. In the steady state, this relation becomes simplified so that $F$ is a linear function of $\log v$.

Studies attempting to reveal the microscopic mechanisms have also been performed following these empirical and phenomenological results. Many types of friction caused by various factors, such as lattice vibrations and electron motion, have been considered in these studies\cite{MDK94,DAK98,MK06,PBFMBMV10,KGGMRM94}. Magnetic friction, which is the frictional force caused by magnetic interactions between spin variables, is one such factor that has attracted attention\cite{WYKHBW12,CWLSJ16,LG18}, and many statistical mechanical models have been proposed\cite{KHW08, Hucht09, AHW12, HA12, IPT11, Hilhorst11, LP16, Sugimoto19, FWN08, DD10, MBWN09, MBWN11, MAHW11}.

In these models, the important behaviors of the system such as the $F$--$v$ relation, differ depending on the choice of model. For example, the Amontons--Coulomb law is observed in some models~\cite{KHW08, Hucht09, AHW12,HA12}, the Stokes law~\cite{MBWN09,MBWN11} in others, while a crossover between these two laws is found in yet other models~\cite{MAHW11}. In our previous studies, we introduced a model that exhibits a crossover or transition from the Dieterich--Ruina law to the Stokes law regardless of whether the range of the magnetic interaction is short~\cite{Komatsu19} or infinite~\cite{Komatsu20}. These studies mainly focused on the steady state, and the behavior of these models in the non-steady state is virtually unknown.

In this study, we introduce a model similar to our previous model\cite{Komatsu19}, and investigate the relaxation of the system when the relative velocity $v$ is changed suddenly  as an example of the non-steady state. Note that in our previous model, the constant external force imposed on the system causes lattice motion, while most other previous studies keep the lattice velocity fixed. Using these dynamics, we successfully described the disturbance of the motion by the magnetic structure, which is why the system exhibited the Dieterich--Ruina law in our previous model. However, to investigate relaxation under a sudden change of $v$, we need to introduce a model where $v$ is given. We therefore introduce a spring connected to the system such that the free end of the spring is pulled at a constant velocity $v$, like in the classical model of friction introduced by Prandtl and Tomlinson\cite{PG12, SDG01, Mueser11}.
In this paper, we introduce the model and its dynamics in Sec.~\ref{Model}, investigate the behavior of the model by numerical simulation in Sec.~\ref{Simulation}, and summarize the study in Sec.~\ref{Summary}.

\section{Model \label{Model} }

\begin{figure}[thbp!]
\begin{center}
\includegraphics[width = 8.0cm]{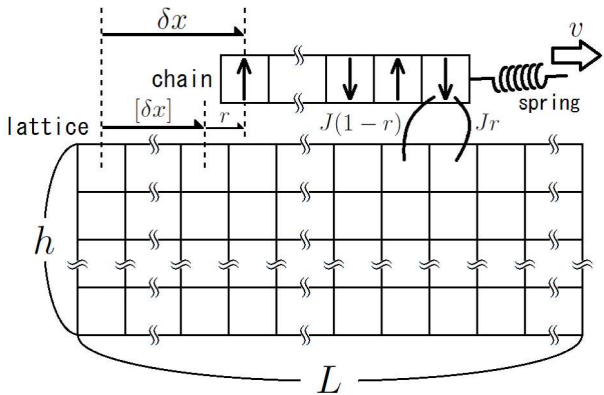}
\end{center}
\caption{ Arrangement of the considered system.  }
\label{settings}
\end{figure}

We consider a chain of length $a$ and a square lattice of side length $L$ and depth $h$ ($a < L$). The chain moves across the upper surface of the lattice as shown in Fig. \ref{settings}, and each lattice point of the chain and the lattice has Ising spins $\left\{ s_n \right\} $ and $\left\{ \sigma _{(i_x , i_y )} \right\} $, where $1 \leq n \leq a$, $1 \leq i_x \leq L$, and $1 \leq i_y \leq h$. We denote the distance the upper chain has moved by $\delta x$. Spins in the lattice interact with each other by nearest-neighbor antiferromagnetic interactions. To simplify the simulation and discussion, we fix the spins in the chain $s_n $ as $s_n = (-1)^{n-1} $, and assume that elastic deformation of the chain and the lattice can be ignored. 
The Hamiltonian of the system is given by:
\begin{eqnarray}
{\cal H}  &= &  J \sum _{\left< \vector{i} , \vector{j} \right> } \sigma _{ \vector{i} } \sigma _{ \vector{j} } \nonumber \\  & + & J \sum _{n=1} ^{a}  \left\{ (1-r) \sigma _{ \left( n+ \left[ \delta x \right] , h \right) } + r\sigma _{ \left( n+ \left[ \delta x \right] +1, h \right) } \right\}  s _{n} , \nonumber \\
\label{Hamiltonian}
\end{eqnarray}
where $\sum _{\left< \vector{i} , \vector{j} \right>}$ is the sum over all pairs of nearest-neighbor spins in the lattice, $\left[ \delta x \right]$ is the largest integer less than or equal to $\delta x$, and $r \equiv \delta x - \left[ \delta x \right]$ is the fractional part of $\delta x$. 
In this Hamiltonian, the $n$th spin in the chain interacts with two adjacent spins in the lattice, $\sigma _{n+ \left[ \delta x \right] }$ and $\sigma _{n+ \left[ \delta x \right] +1 }$, and the coupling constant between them is given by a piecewise linear function of $\delta x$. The dependence of the coupling constant on $\delta x$ is the same as the model in our previous study\cite{Komatsu19}.
To introduce the dynamics of this system, we update the spin variables $\sigma _i$ using the Monte Carlo method, and define the unit time as 1 Monte Caro step (MCS), namely $Lh$ steps of updating. Note that these dynamics have two time scales corresponding to the chain motion and the spin relaxation. We introduce a constant $p_0 (<1)$ and let the acceptance ratio of the update of spins be $p_0$ times that of the normal Metropolis method, namely, $p_0 \min (1, e^{-\beta \Delta {\cal H} } )$, to change the latter time scale.

The chain is attached to a spring with a spring constant of $ak$, and the other end of the spring moves at a constant velocity $v$ such that the chain is pulled by this spring like in the Prandtl--Tomlinson model\cite{PG12, SDG01, Mueser11}. Under these conditions, the chain motion is thought to be dependent on the spring constant. That is, when $k$ is large, even a small extension of the spring cancels the force generated by the magnetic interaction, and the chain is not trapped by the magnetic structure. However, when $k$ is small, we expect the chain to be trapped by this structure. As $k$ decreases, the spring needs to be stretched further to move the chain, meaning that the magnitude of this stretch overwhelms the fluctuation of itself in this case. This means that the fluctuation of the elastic force can also be ignored. Hence, the value of the elastic force is thought to be nearly constant under extremely small $k$.

We let the chain obey the overdamped Langevin equation under a given temperature $T$, so the time development of the shift of the upper chain $\delta x$ is given by
\begin{equation}
0 = - \gamma a \frac{d(\delta x)}{dt} + F_{\mathrm{ex} } - \frac{\partial {\cal H} }{\partial (\delta x)} + \sqrt{2 \gamma T a} R (t) .
\label{Langevin0}
\end{equation}
where $R$ is white Gaussian noise that satisfies $\left< R(t) R(t') \right> = \delta (t-t')$, and the term $\sqrt{2 \gamma T a} R$ is the sum of all of the random forces imposed on the $a$ spin variables in the chain. We adjust the unit of temperature so that the Boltzmann constant $k_B$ is normalized to one. The external force term $F_{\mathrm{ex} }$ is the elastic force from the spring,
\begin{equation}
F_{\mathrm{ex} } = ak(x_{ \mathrm{sp} } (t) - \delta x) ,
\label{Fex_PT}
\end{equation}
\begin{equation}
\mathrm{where}  \ \ x_{ \mathrm{sp} } (t) = vt +x_{ \mathrm{sp}, 0} .
\label{Fex_PT_xsp}
\end{equation}

Substituting these equations into Eq. (\ref{Langevin0}), we get
\begin{equation}
\frac{d(\delta x)}{dt} = \frac{f_{\mathrm{ex} } }{ \gamma } + \frac{1}{ \gamma a } \left( - \frac{\partial {\cal H} }{\partial (\delta x)} + \sqrt{2 \gamma T a} R (t) \right) ,
\label{Langevin1}
\end{equation}
\begin{equation}
\mathrm{where} \ \ f_{\mathrm{ex} } \equiv \frac{F_{\mathrm{ex} } }{a} = k( x_{ \mathrm{sp} } (t) - \delta x) 
\label{fex}
\end{equation}
Note that the frictional force balances with $F_{\mathrm{ex} }$.
To see the relation between the chain motion and the magnetic structure, we also calculate the N\'{e}el magnetization of the part of the lattice contiguous to the chain:
\begin{equation}
n_{\mathrm{touch} } \equiv \frac{1}{a+1} \sum _{n=1} ^{a+1} (-1) ^n \sigma _{ \left( n+ \left[ \delta x \right] , h \right) } .
\label{Ntouch}
\end{equation}

\section{Simulation \label{Simulation}}

In this section, we investigate the friction behavior by numerical simulation. Updating of $\delta x$ is performed after every $\Delta t$ MCSs($=L h \Delta t$ steps) by applying the stochastic Heun method to Eq.~(\ref{Langevin1}). In the actual calculation, the system size is fixed at $L=400, h=40$, and $a=40$, and the other parameters $\gamma, J,$ and $\Delta t$ are given as $\gamma=1, J=1$, and $\Delta t = 0.01$. The constant $p_0$, which is introduced in order to change the acceptance ratio of updating the spin, is fixed at $p_0 = 0.1$. We impose periodic boundary condition in the $x$-direction and open boundary condition in the $y$ direction. 
 First, we calculate the dependences of $f_{\mathrm{ex} } $ and $n_{\mathrm{touch} }$ on the relative velocity $v$ in the steady state. In this calculation, physical quantities are measured over $2 \times 10^5 \leq t \leq 1 \times 10^6$, and averaged over 48 trials independent of each other. The initial state is given as the perfectly antiferromagnetic state with $\delta x = 0 , x_{ \mathrm{sp}, 0} = \frac{2}{k} $. If the initial value of the strain of the spring $x_{\mathrm{sp}, 0} $ is small, the spring requires a longer time to extend especially in the small-$v$ domain. Hence, we let this initial value be larger than the typical value of the strain, $\frac{1}{ak} \left| \frac{\partial {\cal H} }{\partial (\delta x)} \right| \leq \frac{2J}{k} $.

\begin{figure}[hbp!]
\begin{center}
\begin{minipage}{0.99\hsize}
\includegraphics[width = 8.0cm]{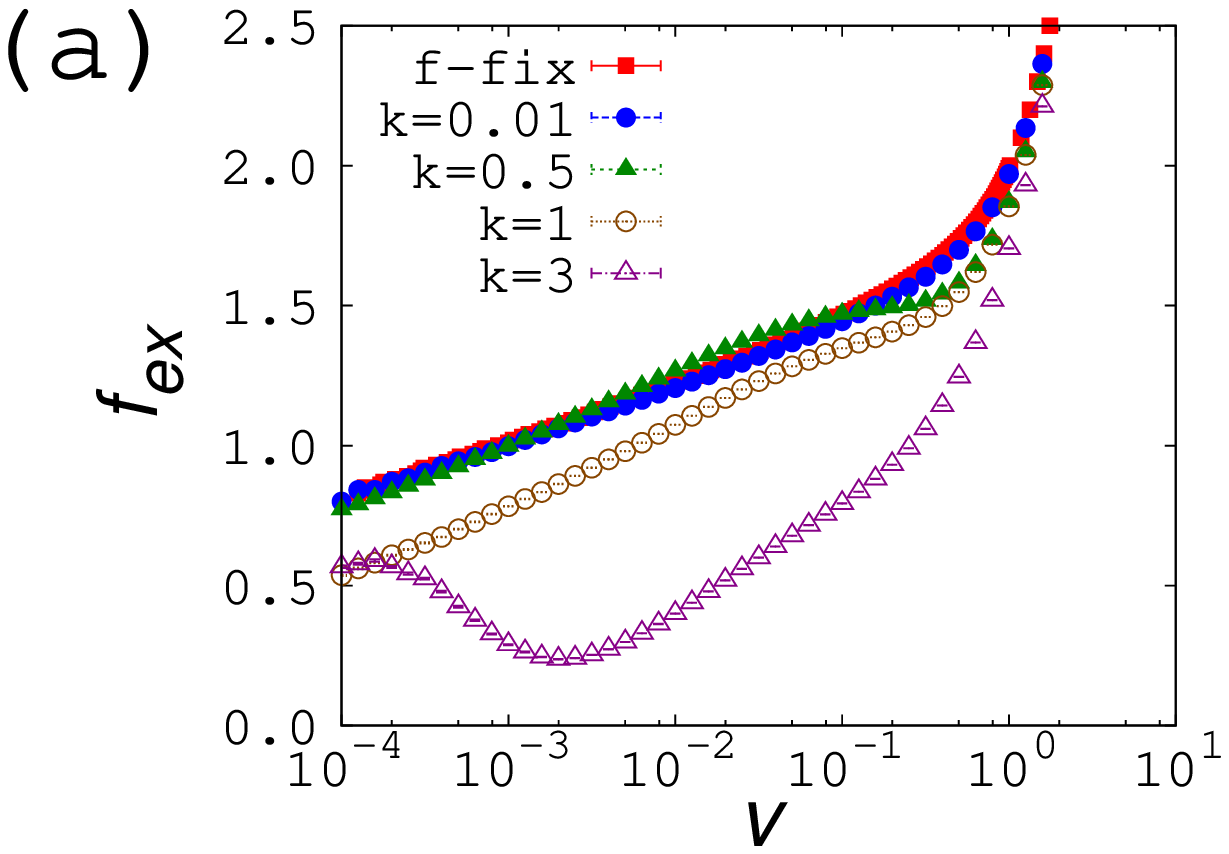} \\ 
\includegraphics[width = 8.0cm]{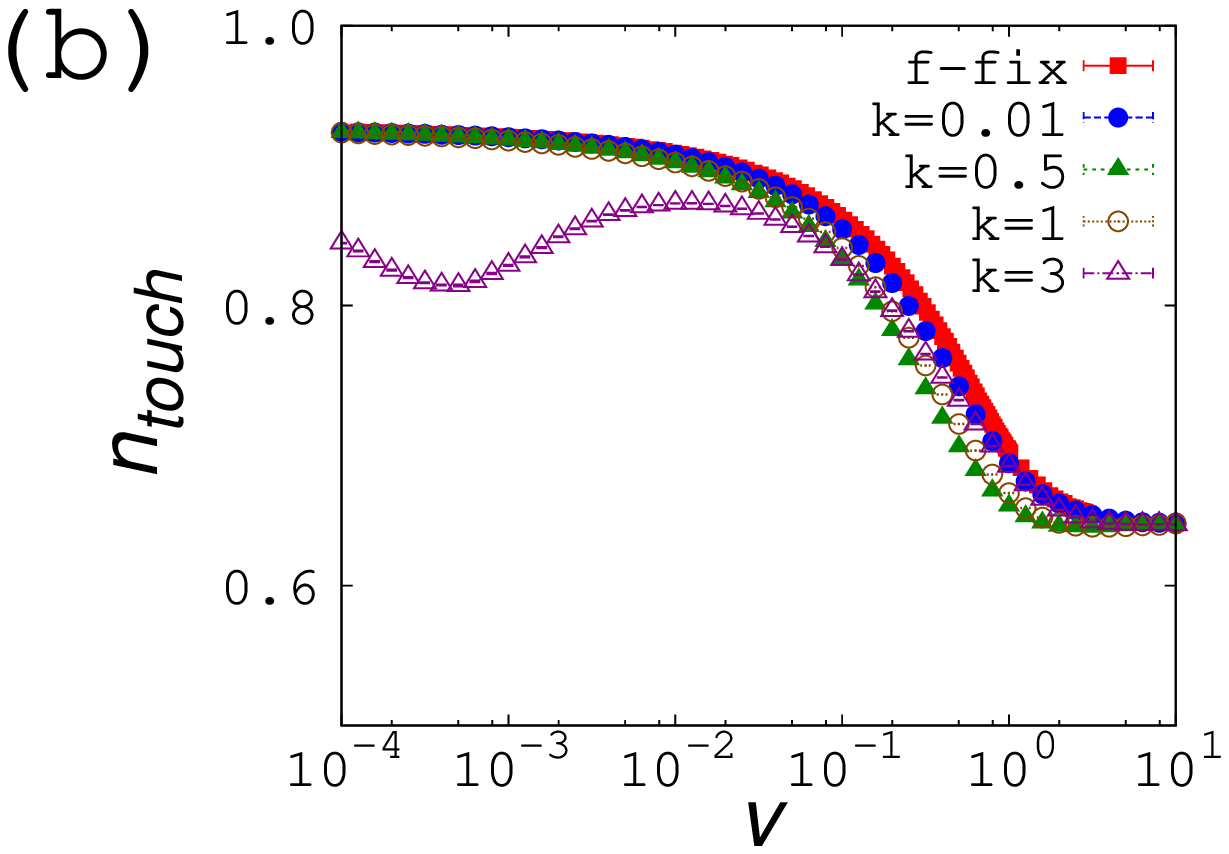} \\ 
\end{minipage}
\end{center}
\caption{(Color online) (a)The $f_{\mathrm{ex} }$-$v$ and (b)the $n_{\mathrm{touch} }$-$v$ relations in the steady state. Red closed squares indicate the case where $f_{\mathrm{ex} } = \mathrm{const.}$, and blue closed circles, green closed triangles, brown open circles, and purple open triangles represent the cases where $f_{\mathrm{ex} } $ is given by Eq. (\ref{fex}) with $k=0.01$, 0.5, 1, and 3, respectively.  }
\label{Fv1}
\end{figure}
For comparison, we also investigate the case where the chain is not pulled by a spring but is driven by a constant external force, $f_{\mathrm{ex} } = \mathrm{const.}$. In the following, we refer to this case as ``$f$-fixed case'' to distinguish it from the normal case in which $f_{\mathrm{ex} } $ is given by Eq. (\ref{fex}).
The result at $T=2.0$ is shown in Fig. \ref{Fv1}. This temperature is lower than the equilibrium transition temperature of the two-dimensional antiferromagnetic Ising model $T_c \simeq 2.27$. We actually also performed the same calculation at $T=2.5 (>T_c)$, but there was no qualitative difference as far as we could find. From Fig. \ref{Fv1}, the $f_{\mathrm{ex} }$-$v$ relation in our model becomes more similar to the $f$-fixed case as $k$ decreases, as we expected in the previous section. Hence, to make the chain motion driven by Eq. (\ref{fex}) similar to that of the ``$f$-fixed case'', we need to make the spring constant $k$ as small as possible. Note that the graph of the $f$-fixed case shows a crossover from the Dieterich--Ruina law to the Stokes law like in our previous models. 

 We next investigate the relaxation process. In this calculation, we let the velocity of the spring $v = v_1$ for $0 \leq t \leq t_W \equiv 2 \times 10^5$, and then change it to $v= v_2$ at $t=t_W$, and measure the time developments of $f_{\mathrm{ex} } $ and $n_{\mathrm{touch} }$. These quantities are averaged over 48000 independent trials. The initial state is almost the same as that of the calculation of the steady state, except that $x_{ \mathrm{sp}, 0} = \frac{2}{k} + x_r $. The parameter $x_r$ is a uniform random value satisfying $0 \leq x_r \leq 2$. If $x_r$ does not exist, $x_{ \mathrm{sp} } (t) $ has a specific value at each time because of Eq. (\ref{Fex_PT_xsp}), and the result of the simulation is biased. Hence, $x_r$ is introduced in order to avoid this problem. 
 Note that in this calculation, the two time scales we mentioned in the previous section, that is those of the relaxations of the spring length and the magnetic structure, have an important role. 

In the case of regular solid surfaces, the frictional force exhibits two relaxation processes, which are a sudden jump and slow relaxation\cite{BC06, DK94}. The latter process is thought to be caused by a slow increase in the contact area accompanying creep deformation. The contribution of this effect is expressed as $\theta$ in Eq. (\ref{DRlaw_theta}). To compare our model with regular solids, we consider the case in which the relaxation of the magnetic structure is sufficiently slower than that of the spring length, similar to creep deformation. 
Letting the value of $f_{\mathrm{ex} }$ in the steady state for a given velocity $v$ be $f_{\mathrm{ex, s} } (v)$, the change in the spring length during the change of velocity can be expressed as
\begin{equation}
\frac{ \left| f_{\mathrm{ex, s} } (v_2 ) - f_{\mathrm{ex, s} } (v_1 ) \right| }{k} .
\end{equation}
The time scale of the relaxation of the spring length $\tau_{\mathrm{sp} }$ can be estimated as the time required to move this length, 
\begin{equation}
\tau_{\mathrm{sp} } \sim \frac{ \left| f_{\mathrm{ex, s} } (v_2 ) - f_{\mathrm{ex, s} } (v_1 ) \right| }{k v_2 } \propto \frac{1}{k v_2 } .
\end{equation}
To make the time scale of the relaxation of the magnetic structure slower than this value, $ \frac{1}{k v_2 }$ needs to be sufficiently small. This means that sufficiently large values of $k$ and $v_2$ need to be chosen. However, as we saw in the calculation of the steady state, if $k$ is too large, the behavior of the system is apparently different from that of the $f$-fixed case and the discussion becomes complicated. Hence, we need to choose a moderate value of $k$ that is not too large or too small. In this calculation, we let $k=0.5$. 

\begin{figure}[hbp!]
\begin{center}
\begin{minipage}{0.99\hsize}
\includegraphics[width = 8.0cm]{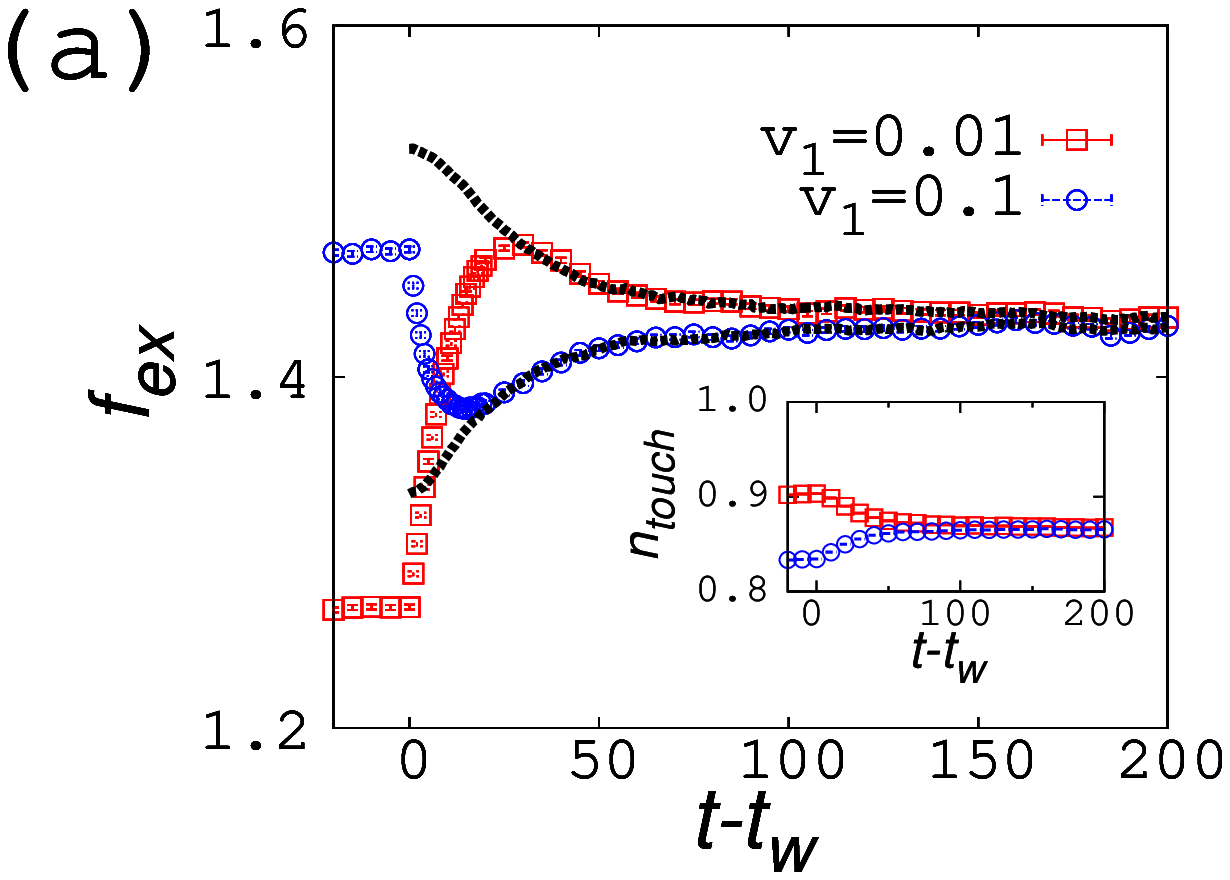} \\ 
\includegraphics[width = 8.0cm]{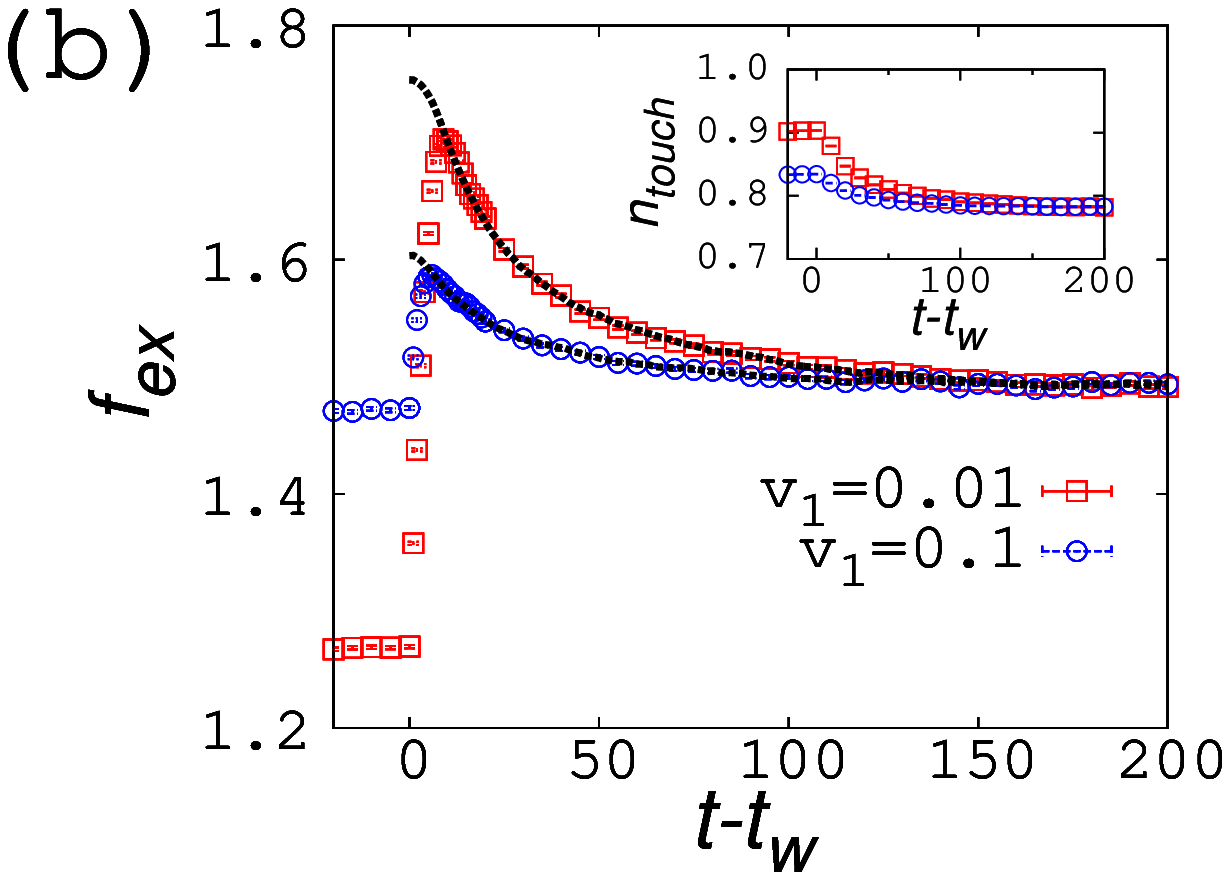} \\ 
\includegraphics[width = 8.0cm]{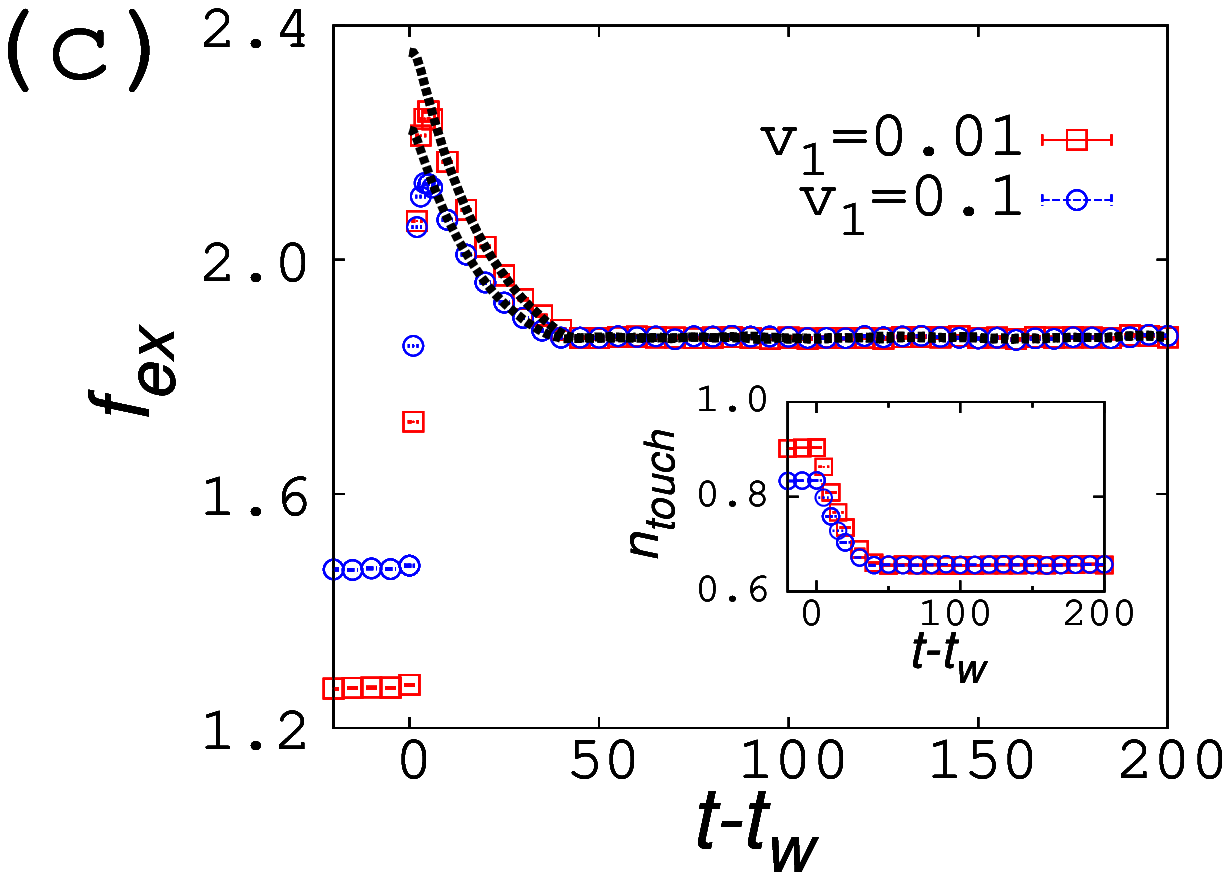} 
\end{minipage}
\end{center}
\caption{(Color online) Time development of $f_{\mathrm{ex} }$ at $k=0.5$ and $v_2 =$ (a) $0.05$, (b) 0.2, and (c) 1.0. Red squares and the blue circles indicate $v_1=0.01$ and 0.1, respectively. Corresponding values of $n_{\mathrm{touch} }$ are plotted in the inset. Dashed lines are curves fitted using Eq. (\ref{DRlaw4}), which is explained later. }
\label{fn_t}
\end{figure} 
The results are shown in Fig. \ref{fn_t}. These graphs shows that the relaxation of $f_{\mathrm{ex} } $ is divided into two processes, a sudden jump and a slow relaxation, like the case of regular solid surfaces. The former process is caused by the fast relaxation of the spring length. Hence, as we discussed above, this needs more time as $v_2$ decreases. Comparing the graphs of $f_{\mathrm{ex} } $ and $n_{\mathrm{touch} }$ in Fig. \ref{fn_t}, the latter process seems to be caused by relaxation of the magnetic structure. We therefore next discuss the relation between $f_{\mathrm{ex} } $ and $n_{\mathrm{touch} }$.

\begin{figure}[hbp!]
\begin{center}
\begin{minipage}{0.99\hsize}
\includegraphics[width = 8.0cm]{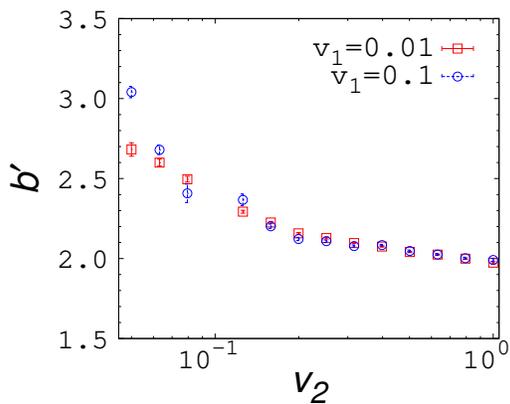} 
\end{minipage}
\end{center}
\caption{(Color online) Dependence of the fitting parameter $b'$ on $v_2$ at $k = 0.5$. Red squares and blue circles are the data at $v_1 = 0.01$, and 0.1, respectively.  }
\label{b1}
\end{figure}

Using a similar discussion to Ref. \cite{Komatsu19}, the velocity $v$ and the external force $F_{\mathrm{ex} }$ are expected to obey the relation
\begin{equation}
\log v = \frac{ \alpha F_{\mathrm{ex} } - U_0 }{T}  + c ,
\label{DRlaw2}
\end{equation}
where $U_0$ is the height of the potential barrier made by the magnetic structure, $\alpha$ is the distance between the top and bottom of the potential barrier, and $c$ is a constant. In our model where the spins in the chain are fixed in a perfectly antiferromagnetic order, the height of the potential is thought to be proportional to $n_{\mathrm{touch} }$. Hence, by using a constant $b$, $U_0$ can be expressed as $U_0 = b n_{\mathrm{touch} } $ and Eq. (\ref{DRlaw2}) can be expressed as
\begin{equation}
\log v = \frac{ \alpha F_{\mathrm{ex} } - b n_{\mathrm{touch} } }{T}  + c .
\end{equation}
Transforming this equation gives
\begin{eqnarray}
f_{\mathrm{ex} } & = & \frac{F_{\mathrm{ex} } }{a} = \frac{1}{a \alpha} \left( T \log v + b n_{\mathrm{touch} } - Tc \right) \nonumber \\
& = & \alpha ' \log v + b' n_{\mathrm{touch} } + c'  ,
\label{DRlaw3}
\end{eqnarray}
\begin{equation}
\mathrm{where} \ \ \alpha ' =  \frac{T}{a \alpha} \ , \ b' =  \frac{b }{a \alpha} \ , \ c' = - \frac{Tc}{a \alpha} .
\end{equation}
According to Eq. (\ref{DRlaw3}), $f_{\mathrm{ex} } $ is a linear function of $n_{\mathrm{touch} } $, and $v$ depends on the magnetic structure through this relation. When the values of $f_{\mathrm{ex} } $ and $n_{\mathrm{touch} } $ in the steady state are already known as $f_{\mathrm{ex, s} } $ and $n_{\mathrm{touch, s} } $, Eq. (\ref{DRlaw3}) can be rewritten as
\begin{equation}
f_{\mathrm{ex} } = f_{\mathrm{ex, s} } + b' \left( n_{\mathrm{touch} }  - n_{\mathrm{touch, s} } \right) .
\label{DRlaw4}
\end{equation}
Note that the above discussion regarding the height of the potential is simplified by the fixed spins in the chain. If these spins were not fixed, the contribution of the magnetic structure to the frictional force, which appears in Eqs. (\ref{DRlaw3}) and (\ref{DRlaw4}) as the term $b' n_{\mathrm{touch} } $, would be more complicated. To examine whether the time-development data in Fig. \ref{fn_t} actually obey Eq. (\ref{DRlaw4}), we fit the values of $f_{\mathrm{ex} } $ and $n_{\mathrm{touch} } $ at each time to Eq. (\ref{DRlaw4}) by the least-squares method. We use the data satisfying $25 \leq t - t_W < 80$ when $0.05 \leq v_2 < 0.1$, and those satisfying $15 \leq t - t_W < 80$ when $0.1 < v_2$, and we impose the same weight at every point. The values of $f_{\mathrm{ex, s} } $ and $n_{\mathrm{touch, s} } $ are taken from the data in Fig. \ref{Fv1}. The results are plotted as the dashed lines in Fig. \ref{fn_t}. From these figures, since the fitting curves seem to reproduce the slow-changing part of the relaxation, the above discussion is thought to be correct. However, since the value of $b'$ changes depending on $v_1$ and $v_2$, we performed similar calculations for several values of $v_2$, and visualized this dependence as Fig. \ref{b1}.
This figure shows that $b'$ has a nearly constant value in the large-$v_2$ domain, and becomes larger when $v_2$ gets smaller. Note that relaxation of the spring length is slow when $k$ is small, as we can see in Fig. \ref{fn_t}(a). Hence, there is a possibility that the data fittings in the small-$v_2$ domain are inaccurate because of the incomplete relaxation of the spring.  We also note that the above discussion deriving Eqs. (\ref{DRlaw3}) and (\ref{DRlaw4}) assumes that the external force $F_{\mathrm{ex} }$ is nearly constant during the chain motion, and the detachment from this assumption is also thought to cause the non-constant $b'$. These points make the discussion of the behavior of $b'$ within the range of our calculation difficult. 
 
\begin{figure}[hbp!]
\begin{center}
\begin{minipage}{0.99\hsize}
\includegraphics[width = 8.0cm]{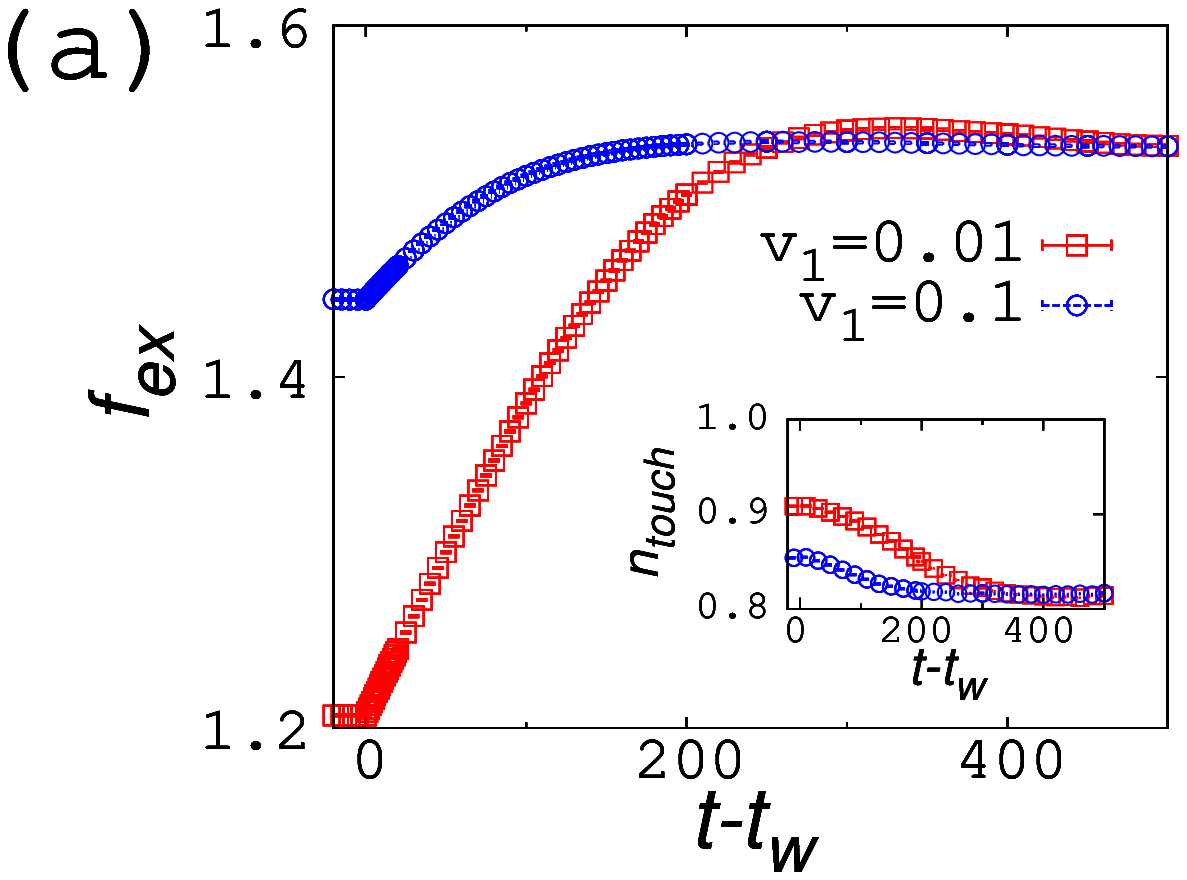} \\ 
\includegraphics[width = 8.0cm]{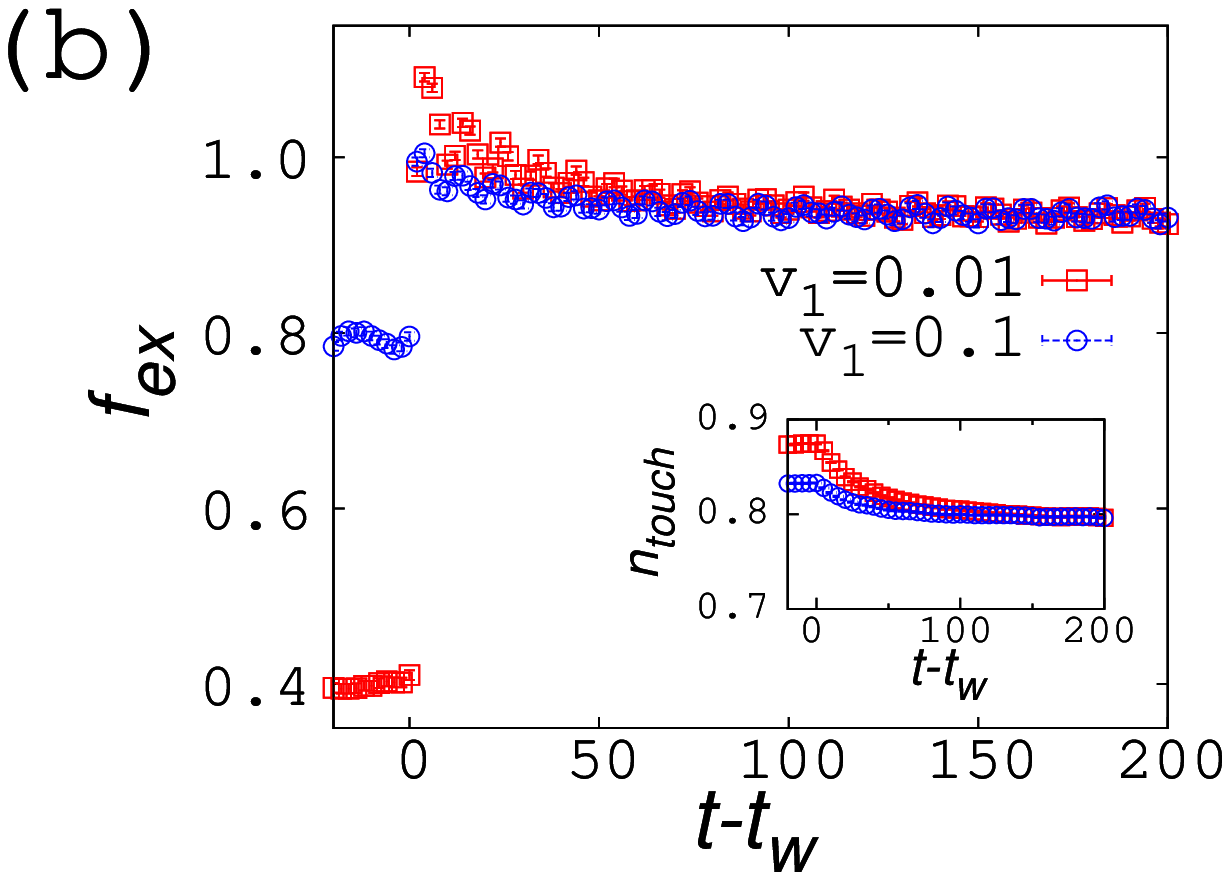}  
\end{minipage}
\end{center}
\caption{(Color online) Time development of $f_{\mathrm{ex} }$ at $v_2 = 0.2$. The spring constant $k$ is given as $k=$ (a) $0.01$, and (b) 3. The meanings of the points are the same as Fig. \ref{fn_t}. Corresponding values of $n_{\mathrm{touch} }$ are plotted in the inset. }
\label{fn_tk}
\end{figure}

Finally, we calculate the relaxation process when $k$ is small or large, keeping other parameters and conditions the same as that of Fig. \ref{fn_t}. The results at $k=0.01$ and 3 are shown in Figs. \ref{fn_tk} (a) and (b), respectively. From Fig. \ref{fn_tk}(a), the relaxation of the spring is slow and the distinction between the two relaxation processes is ambiguous when $k$ is small. 
In the case of large $k$, the chain is not trapped by the magnetic structure as we discussed in Sec. \ref{Model}. The force generated by the magnetic interaction itself is proportional to $n_{\mathrm{touch} } $ even in this case, so the dependence of $f_{\mathrm{ex} } $ on $n_{\mathrm{touch} } $ remains and the two processes of the relaxation can be also observed like in Fig. \ref{fn_tk}(b). However, the value of $f_{\mathrm{ex} } $ in this case is susceptible to slight changes in $\delta x$ during the chain motion. As a result of this subtleness, the time development of $f_{\mathrm{ex} } $ contains an oscillation that cannot be explained by Eq. (\ref{DRlaw4}).

\section{Summary \label{Summary}}

In this study, we considered a system composed of a chain and a lattice interacting with each other by magnetic interactions, and investigated the behavior of the magnetic friction. By attaching a spring to the chain such that the opposite end of the spring moves at a constant velocity $v$, we calculated the relation between the frictional force and $v$. 
In particular, we investigated the relaxation process of the frictional force and the magnetic structure after a sudden change in $v$. In this calculation, two relaxation processes, namely a sudden jump and a slow relaxation, were observed like the case of regular solid surfaces if the spring constant has a moderate value. The latter process results from relaxation of the magnetic structure in our model, but is caused by creep deformation in the case of regular solid surfaces.
The distinction between these two processes is clear when the time scale of the relaxation of the magnetic structure is sufficiently slower than that of the spring length. When the value of the spring constant is too small, these time scales become comparable with each other and the two processes are not observed.
Note that we can modulate the time scale of the relaxation of the magnetic structure by changing the constant $p_0$. If we slow this time scale by adopting smaller $p_0$, two processes are clearly observed in the case with smaller $k$. The discussion in Sec. \ref{Simulation} deriving Eq.(\ref{DRlaw3}) itself is thought to be more accurate for the case where $k$ has a smaller value, because this discussion assumes that the external force imposed by the spring is nearly constant during the chain motion. Hence, we need to investigate whether the problem of the non-constant fitting parameter $b'$ discussed in Sec. \ref{Simulation} is eliminated in cases with smaller values of $k$ and $p_0$. This investigation requires a large amount of computational time, and is left as future work.

To examine whether our model obeys the Dieterich--Ruina law given by Eqs. (\ref{DRlaw}) and (\ref{DRlaw_theta}) in the non-steady state, we need to investigate the time development of the magnetic structure carefully, and compare it with that of $\log \theta$ in Eq. (\ref{DRlaw}). Since relaxation under a sudden change in $v$ considered in our study is too simple and has insufficient information to complete this investigation, the behavior of magnetic friction under more complicated situations also needs to be studied in the future.

\section*{Acknowledgments}
Part of the numerical calculations were performed on the Numerical Materials Simulator at National Institute for Materials Science.


\end{document}